\documentclass[11pt,oneside,a4paper]{article}

\usepackage{cite}
\usepackage{amsmath,amssymb,amsfonts}
\usepackage{microtype}
\usepackage{graphicx}
\usepackage{textcomp}
\usepackage{xcolor}
\usepackage{hyperref}

\def\BibTeX{{\rm B\kern-.05em{\sc i\kern-.025em b}\kern-.08em
    T\kern-.1667em\lower.7ex\hbox{E}\kern-.125emX}}
\begin{document}

\title{Digital Sovereignty and Software Engineering\\for the IoT-laden, AI/ML-driven Era}

\author{Christian Berger, christian.berger@gu.se}

\maketitle

\begin{abstract}
Today's software engineering already needs to deal with challenges originating from the multidisciplinarity that is required to realize IoT products: Many variants consist of sensor/actuator-powered systems that already today use AI/ML systems to better cope with the unstructuredness of their intended operational design domain (ODD), while, at the same time, such systems need to be monitored, diagnosed, maintained, and evolved using cloud-powered dashboards and data analytics pipelines that process, aggregate, and analyze countless data points preferably in real-time.
This position paper discusses selected aspects related to \emph{Digital Sovereignty} from a software engineering's perspective for the IoT-laden, AI/ML-driven era: While we can undeniably expect more and more benefits from such solutions, a specific light shall be shed in particular on challenges and responsibilities at design- and operation-time that, at minimum, prepare for and enable or, even better, preserve and extend digital sovereignty from a software engineering's perspective.
\end{abstract}


\section{Introduction}
\label{sec:introduction}

Our societies are growingly embracing the benefits and opportunities of Internet-of-Things (IoT)- and AI/ML-powered aspects of our lives. Visibly are we knowingly dependent on our smartphones to enable a more and more digitalized lifestyle that is characterized by use cases, which shifted from pure amusement and entertainment like playing games or watching movies to seeing benefits from IoT-enabled digital workflows for intelligent power grid management or personalized health advises powered by intelligent wearables for example. However, also invisibly did our societies enrich our daily life with IoT-powered devices or cyber-physical systems to replace human workforce in logistics centers by using robot-powered shelves or monitor waste water flows in pipes in metropolitan areas to better plan maintenance activities. We are more and more enjoying the benefits of IoT-devices and cyber-physical systems powered by growingly complex AI/ML-systems, and can expect an even more prosperous future with further applications like self-driving vehicles, driverless goods logistics even for the ``last mile'', robotic assistance for care-taking, and further application scenarios that benefit from connectivity (cf.~Giaimo et al., \cite{GIAIMO2020110781}, Mallozzi et al., \cite{mallozzi2019autonomous}).

\section{Today's Expectations on\\Full-Stack IoT-Software Engineers}
\label{sec:today}
Today's software engineers are part of interdisciplinary engineering teams that design, realize, systematically validate and verify, and operate growingly complicated solutions for IoT-systems.

\begin{table}[htbp]
    \centering
    \begin{tabular}{|p{8.4cm}|}
        \hline
        An IoT system can be understood as a cyber-physical system (CPS) that may contain one or multiple sensors and possibly some actuators to sense and interact within its operational design domain (ODD), while also being (continuously) connected with the Internet and hence, having access to computationally and functionally powerful cloud-backends.\\
        \hline
    \end{tabular}
\end{table}

Open vacancy announcements to hire engineers for creating solutions based on the aforementioned definition of IoT also emphasize the capability for the successful candidate to contribute as \emph{full-stack software engineer} who not only masters one aspect of the complete system but preferably all aspects: The user- and IoT-facing cloud front-end, the operational cloud back-end, and the IoT unit itself, ie., the software stack on a sensor or robotic unit that may even contain AI/ML-powered components. Already this high-level description indicates a large variety of methods and tools to:
\begin{enumerate}
    \item Design, implement, train, and validate the AI/ML-components, and
    \item Design, implement, test, package, and deploy a growing amount of microservices in scalable cloud-environments, and
    \item Design, implement, test, package, deploy, and tune embedded software stacks exploiting silicon-accelerated functionalities--oftentimes under computational, connectivity, and memory constraints, and
    \item Designing, monitoring, automating, and accelerating the necessary engineering pipelines (ie.,  \emph{GitOps}) that trigger complex processes:
    \begin{enumerate}
        \item CI/CD pipelines to integrate and test the software change with the distributed software stack, including the preparations for target deployment
        \item ML pipelines that evaluate changes in an ML-model's network architecture, or in the underlying training and validation datasets to calculate changes in key performance indicators (KPIs) for the AI/ML-components
        \item Planning and preparing data-driven experiments that are also known as continuous experimentation (CX), where the population of IoT-unit under monitoring may be split into sub-populations to systematically evaluate the performance of the latest software changes (cf.~Giaimo and Berger, \cite{GB20})
    \end{enumerate}
\end{enumerate}

The aforementioned items clearly indicate that today's software engineering challenges to develop and improve IoT-systems are manifold and in industry oftentimes realized with a broad variety of different tools. Using domain-specific languages (DSLs) to describe workflow steps on a high abstraction level to feed tools that deal with the lower level technical complexities and that automate the artifact transformation processes are unavoidably necessary. Best practices have in recent years started to emerge even across industries to exchange ideas, concepts, and process metrics like in Google's DORA reports.\footnote{Cf.~\url{https://www.devops-research.com/research.html}}

Latest trends for artifact transformation processes are not only focusing on the data-driven CI/CD aspects but started to also cover growing amounts of the operational environment. Key ideas therein are \emph{infrastructure-as-code} as well as \emph{software-defined X} (where X can be network, storage, computation, \dots) that relate to an increased degree of abstraction to hide hardware intricacies and dependencies. The clear goal is to replace them with \emph{software entities} that can be captured and described in software artifact transformation processes to accomplish an even higher level of end-to-end automation from software change during the \emph{development} phase to deployment into the target platforms for the actual \emph{operations} phase. Companies have realized that fully formalized, automated, and rapid data-driven CI/CD/CX-pipelines behind \emph{GitOps} are a key-differentiating way-of-working that not only sets high performing companies apart from the average (cf.~Google DORA report$^1$), but that also enables innovation potential and capabilities for future product enhancements and business opportunities. 

\section{Upcoming Challenges and Responsibilities for Full-Stack IoT-Software Engineers}
\label{sec:tomorrow}

As we have seen in the introduction in Sec.~\ref{sec:introduction}, IoT-systems and IoT-powered solutions growingly permeate our societies and daily lives in visible and invisible ways so that we are becoming more and more dependent on their uninterrupted and fault-tolerant operation. Hence, software engineers do not only face today's challenges from (a) growing functional complexities motivated by the trend to include AI/ML-components to tackle an ODD's unstructuredness, (b) continuous pressure to embrace adaptability in software and system architectures (cf.~Zavala et al., \cite{zavala2021adaptive}) motivated by scalability expectations for large software stacks that shall exploit silicon-powered computation accelerators, and (c) traceable, data-driven validation and verification from code changes to explanations for unexpected behavioral patterns in a system's ODD.

Software engineers who operate with these aforementioned growingly software and system architectures, as well as their underlying artifact transformation processes face knowingly and unknowingly design decision that may impact and constrain future design possibilities in the on-going society's digitalization transformation. Hence, the awareness of such impact and constraints must be raised--to, at minimum, preserve the space of future design possibilities, or, even better, extend the possibilities and opportunities while striving for fair, inclusive, and equal data-driven decision making in both, products \emph{and} engineering processes.

Sovereignty originates from the Latin word \emph{superanus} via the French word \emph{souverainet\'{e}} and is literally expressing a power relationship\footnote{Cf.~Encyclopedia Britannica, \url{https://www.britannica.com/topic/sovereignty}}. Underlying ideas have lately been carried over to modern times and found their way into the \emph{Berlin Declaration on Digital Society and Value-Based Digital Government}\footnote{Cf.~\url{https://ec.europa.eu/newsroom/dae/document.cfm?doc_id=75984}} that state: ``Digital sovereignty is key in ensuring the ability of citizens and public administrations to make decisions and act in a self-determined manner in the digital world.'' Hence, within software engineering, it can be derived: 

\begin{table}[htbp]
    \centering
    \begin{tabular}{|p{8.4cm}|}
        \hline
        Dealing with digital sovereignty in software engineering relates to near-term decision making for products and processes during a software system's development as well as its operations that may impact or constrain the long-term self-determinism of the decision making for this software system.\\
        \hline
    \end{tabular}
\end{table}

Based on the aforementioned definition, a selection of examples are presented and discussed in the following.

\emph{Open Source} is a very apparent example that comes to mind when thinking about preserving self-determinism of the decision making process for digital goods. A clear aspect for individuals and corporations relates to the decision in near-term what type of open source license shall be used as the variety ranges from permissive to ``copy-left''\footnote{Cf.~\url{https://choosealicense.com}}: Some licenses like MIT allow not only to \emph{take and do} with a 3rd party software whatever one is preferring, while others like GPL requires changes to be shared back with the community. Both scenarios may make sense from an individual's or corporation's strategy but clearly constrain the future self-determinism of the decision making as replacing a 3rd party's software component later is causing potentially unwanted effort.

\emph{Open Data} can be considered as open source software's sibling. EU's Directive 2019/1024 states that ``\dots data is open if it can be freely used, re-used and shared by anyone for any purpose''. Hence, next to the open source licenses for software, there are licenses for data such as Creative Commons\footnote{Cf.~\url{https://creativecommons.org/choose/}} that follow a similar spirit as the previous topic.

\emph{Data that is used for training an AI/ML component} is apparently also a potential source for constraining the future self-determinism of the decision making about the own software system as designing unknowingly biased training and validation datasets, as well as lacking an understanding about their representativeness and completeness for working with AI/ML-pipelines: What data is collected and, maybe more important, what data is \emph{not} collected? While these aspects are hard problems to describe, model, and measure, correcting a potentially skewed dataset as well as the AI/ML-powered IoT-unit based on this later during operations is challenging.

\emph{3rd party software components} constitute in many software systems today already a significant part to avoid \emph{re-inventing the wheel} and to allow focusing on the core and differentiating aspects of a software system. However, the inclusion of such 3rd party software can take place in various ways: (a) The external software can simply be linked with the own system as part of a dependency installation, or (b) it can be included as a traceable software asset by code cloning (for instance, copying a header-only C++ library into the own project), or (c) by re-implementing a required functionality. The potentially constraining decision in such cases is related to the \emph{update and release frequency} of the external software and how this shall be incorporated with the own software system: Are there different release channels such as \emph{stable} and \emph{edge} available? What are the long-term project vision and ideas for the external software project? How active, responsive, and inclusive is the developer community around the external project? How are vulnerabilities and bugs handled in general?

\emph{Cloud infrastructure} has already become a differentiating success factor for companies to scale depending on varying workloads for web systems for example. Apparently, outsourcing computing and storage infrastructure to large cloud operators allows individuals and corporations to focus on core competencies for the own software systems on the one hand. However, it is also very clear on the other hand that there is the risk for a \emph{tech lock-in} potentially threatening the self-determinism of the decision making in the long-term when moving from one cloud operator to another or even back to \emph{on premise}. This trend is gaining further relevance with the recent rise of \emph{serverless computing} where even the cloud operators provide the complete management of the resource provisioning and adaptive scaling so that software engineers can primarily focus on a system's functionality without having to worry what particular virtual machine type or storage backend needs to be chosen. While this trend is aligning even with \emph{infrastructure-as-code} or \emph{software-defined X}, the possibilities of a cloud operator and its plans for future adjustments are influenced by its commercial design drivers, which potentially constrain the future self-determinism of the own decision making for the own software system. 

\emph{Development environments} are lately also benefitting from advancements in AI/ML-assisted coding: While modern integrated development environments (IDEs) support code completion based on automatically created and maintained API indexes, recent approaches like GitHub Copilot\footnote{Cf.~\url{https://copilot.github.com}} push this even further to let developers describe in prose (ie., by code comments in plain English) what is needed in the current part of the source code and the ``AI-pair programmer'' is generating the matching and functional code that is adapted to the syntactical context surrounding the code fragment. While this appears to be a helpful initiative at first sight, there are various aspects that potentially threaten future self-determinism of the own decision making for the own software system: (a) On what corpus were such AI-pair programmers trained in terms of algorithmic design (style, performance, legal license), or (b) the increased emphasis to fully understand the auto-generated code fragments that were generated and included into the own code base that needs to be maintained in upcoming refactorings, or (c) the auto-generated code is following a certain implementation design like functional or imperative programming that may be in conflict with the own design, or (d) the auto-generated code may be even non-optimized and for resource-constrained IoT environments and such auto-generated fragments would have to be replaced later to meet realtime constraints.  

\emph{Digital twins} of IoT-units are considered a promising cloud-supported companion to monitor and diagnose a physical unit such as an IoT-system. While such digital twins allow for data analysis by using data collected from the past to be used for AI/ML components, current ideas also consider digital twins to be part of the functionality of an IoT-unit to, for instance, conduct ``what-if'' analyses like cloud-based route planning for self-driving vehicles by using traffic flow information from other self-driving vehicles nearby. The underlying models for such digital twins are inherently based either on data collected in the past or by being based on assumptions of the systems and their operational context. While the former is constrained by the design decision for its perception system to ``understand'' an IoT's surrounding, the latter may apparently be unknowningly based on false assumptions at design time and hence, potentially limiting the future self-determinism of the own decision making.

\section{Conclusions and Future Research}
In this position paper, today's expectations for software engineers working with AI/ML-powered IoT-units were described and discussed. Based on the already demanding expectation on such full-stack IoT-software engineers, selected upcoming challenges and responsibilities for these full-stack IoT-software engineers were presented. These examples underlined that awareness for dealing with digital sovereignty in software engineering must be raised as near-term decision making for these connected, cyber-physical systems during their development as well as later during their operations may impact or constrain the long-term self-determinism of the decision making for such IoT-systems. Hence, future research must focus on methods to explicate this responsibility for digital sovereignty in software system design and operations.

\bibliography{references} 

\end{document}